\documentclass[%
 reprint,
superscriptaddress,
amsmath,amssymb,
aps,
floatfix,
longbibliography,
]{revtex4-2}

\usepackage{graphicx}
\usepackage{dcolumn}
\usepackage{bm}
\usepackage{booktabs}
\usepackage{graphicx}
\usepackage{dcolumn}
\usepackage{bm}
\usepackage{nowidow}
\usepackage{tabularx}
\usepackage{siunitx}
\usepackage[usenames,dvipsnames]{xcolor}
\usepackage[colorlinks=true, allcolors=black]{hyperref}
\usepackage{tikz}
\usepackage{times}
\usepackage{hyperref}
\usepackage{afterpage}
\usepackage{float}
\hypersetup{
    citecolor = MidnightBlue,
    linkcolor = RedOrange
}
\usepackage{natbib}
\setcitestyle{square, comma, numbers,sort&compress, super}
\begin{document}

\newcommand{\YF}[1]{\textcolor{purple}{\,#1}}
\newcommand{\RS}[1]{\textcolor{magenta}{\,#1}}
\newcommand{\LI}[1]{\textcolor{olive}{\,#1}}
\newcommand{\SJ}[1]{\textcolor{blue}{\,#1}}
\newcommand{\ED}[1]{{\color{cyan}#1}}


\title{Extracting hydrogel properties by watching hydrogel particles moving through solid ice}

\author{Yanxia Feng}
\affiliation{Department of Materials, ETH Z\"{u}rich, 8093 Z\"{u}rich, Switzerland.}%

\author{Se-Hyeong Jung}
\affiliation{Department of Materials, ETH Z\"{u}rich, 8093 Z\"{u}rich, Switzerland.}%

\author{Camillo Sirvinski}
\affiliation{Department of Materials, ETH Z\"{u}rich, 8093 Z\"{u}rich, Switzerland.}%

\author{Dan Balkanyi}
\affiliation{Department of Materials, ETH Z\"{u}rich, 8093 Z\"{u}rich, Switzerland.}%

\author{Federico Paratore}
\affiliation{Department of Materials, ETH Z\"{u}rich, 8093 Z\"{u}rich, Switzerland.}

\author{Lucio Isa}
\affiliation{Department of Materials, ETH Z\"{u}rich, 8093 Z\"{u}rich, Switzerland.}%

\author{Robert W. Style}
\email[]{robert.style@mat.ethz.ch}
\affiliation{Department of Materials, ETH Z\"{u}rich, 8093 Z\"{u}rich, Switzerland.}%

\date{\today}

\begin{abstract}
Strikingly, when hydrogel particles are embedded in ice in a temperature gradient, they move through the solid ice towards warmer temperatures, while swelling as they warm up.
This motion comes from a flow of unfrozen water through the hydrogel mesh, driven by a process known as `cryosuction'.
Here, we show how one can use this behavior to measure -- with extremely high resolution -- a range of different hydrogel transport properties, and how these change as a hydrogel deswells.
These properties include permeability, compressibility, and poroelastic diffusivity: all of which are challenging to measure, but widely important for phenomena involving swelling, dehydration, transpiration and filtration.
We demonstrate the measurement technique using poly(ethylene glycol) diacrylate (PEGDA) hydrogels.
The technique uses picoliter-scale hydrogel volumes, and yields measurements of hydrogel properties that are consistent with existing literature data.
The high resolution of our measurements also allows us to test commonly-used, classical predictions for hydrogel properties (derived assuming an idealized, homogeneous polymer network in the hydrogel).
We show that these classical predictions do not work well, highlighting the need for new models that can accurately describe real hydrogels.

\end{abstract}

\maketitle

Hydrogels are versatile class of materials, which have found widespread use -- often due to their ability to dramatically swell and shrink, or due to their high water permeability.
For example, drug-loaded hydrogels swell in the body and release the drugs at controlled rates \cite{kim1992hydrogels,li2016designing},
superabsorbent hydrogels rapidly absorb spilled liquids \cite{kabiri2011superabsorbent},
plants and swelling-induced actuators move or develop forces by swelling \cite{ionov2013biomimetic,hofhuis2016morphomechanical,shen2026programmable}, and
lubrication in articular cartilage changes as fluid is squeezed out of pressurized joints \cite{ateshian2009role}.
Other common examples arise in fields ranging from agriculture \cite{kaur2023hydrogels}, to food science \cite{li2021hydrogel}, soft robotics \cite{lee2020hydrogel}, tissue engineering \cite{tibbitt2009hydrogels}, membrane science \cite{bell2005biomedical,eddine2024tuning}, and personal care \cite{bashari2018cellulose}.

Because water transport is so fundamental to such hydrogel behavior, we need accurate techniques to measure the key hydrogel transport properties.
These properties include permeability ($k$), compressional stiffness ($B$), and the poroelastic (or cooperative) diffusivity ($D$) \cite{hu2010using,kopecz2023mechanical,sakai2020physics}.
However, these are very hard to measure -- largely due to how slowly water moves through typical gels \cite{grattoni2001rheology, kapur1996hydrodynamic, white1960permeability, johnson1996hydraulic, ju2010characterization}.
For example, to measure $k$, we would need to measure the flux of water, $\mathbf{Q}$, driven through a hydrogel membrane by a pressure gradient, $\mathbf{\nabla p}$. This is related to $k$ via Darcy's law:
\cite{hegde2022two,webber2023linear}:
\begin{equation}
    \mathbf{Q}=-\frac{k}{\eta} \mathbf{\nabla p},
    \label{eqn:darcy}
\end{equation}
where 
$\eta$ is water's dynamic viscosity ($\eta=8.9\times10^{-4}\,$Pa.s at 25$\,^\circ$C).
In typical hydrogels, $k\sim 10^{-18}\, \mathrm{m}^2$ \cite{gao2022quantifying}.
Then, a 1kPa pressure difference across a 1mm-thick membrane would give $\mathbf{\nabla p}=10\,$MPa/m, driving an extremely slow flux of $Q \sim 1\,$nm/s -- which is hard to detect.
Measuring $B$ also requires lengthy experiments, where hydrogels are equilibrated under a series of compressive loads \cite{gao2021scaling}.
At each different load, we must wait for excess water to drain out of the hydrogel, which takes a time $t_D\sim h^2/D$, where $h$ is the sample thickness.
$D\sim 10^{-10}\,\mathrm{m}^2/\mathrm{s}$ in typical hydrogels \cite{caccavo2018hydrogels}.
Thus, a 1\,mm-thick sample will take $10^4\,\mathrm{s}\approx 3\,$hr to equilibrate at every data point, making measurements extremely time consuming.
As a result of such challenges, we lack high precision measurements of water transport properties in hydrogels, especially measurements that show how these properties vary with hydrogel swelling.

Here, we demonstrate a technique to precisely measure hydrogel transport properties.
Small hydrogel particles are embedded in ice, and exposed to a temperature gradient. Strikingly, the particles then migrate through the ice towards warmer temperatures, while slowly swelling. By measuring the speed and size of the particles as a function of temperature, we extract the transport properties with excellent resolution.

\subsection{Experimental set-up}

To showcase this technique, we freeze poly(ethylene glycol) diacrylate (PEGDA) hydrogel particles in a custom-built freezing apparatus.
We create the particles via droplet-based microfluidics, using PEGDA with molecular weight of 700 g/mol, and with a polymer content of $\phi_p=20$ vol\% (see Materials and Methods).
Particles are purified and re-dispersed in deionized (Milli-Q) water, after which they have radii between 8-14 $\mu$m.
N.B. we choose to work with 20\% PEGDA gels, as these gels do not swell from their as-prepared state when re-dispersed in water (see \cite{feng2025characterizing,feng2025controlling}).

We freeze dilute hydrogel-particle suspensions in a thin sample cell in the apparatus shown in Figure \ref{fig:schem_isothermal}A \cite{gerber2022stress,gerber_stage} (further details in Materials and Methods).
This essentially consists of two sets of copper blocks which sandwich the two ends of the sample cell, leaving a 2\,mm gap for visualizing samples (see Figure \ref{fig:schem_isothermal}B).
We can set the temperatures of the two sets of blocks, $T_1,T_2$, to temperatures between  -25 and 25$^\circ$C.
When $T_1=T_2<0\,^\circ\mathrm{C}$, we freeze the whole sample isothermally. 
When $T_1<0\,^\circ\mathrm{C}<T_2$, we make a temperature gradient, $G$, in the sample cell (see SI), with a freezing front that is visible in the viewing gap.
We measure the precise temperature field with thermistors in the cell, and by using the ice-water interface as a reference point with $T=0\,^\circ\mathrm{C}$ (see SI for details).

\subsection{Isothermal freezing of hydrogel particles in ice}

\begin{figure}[hpbt]
    \centering
    \includegraphics[width=\linewidth]{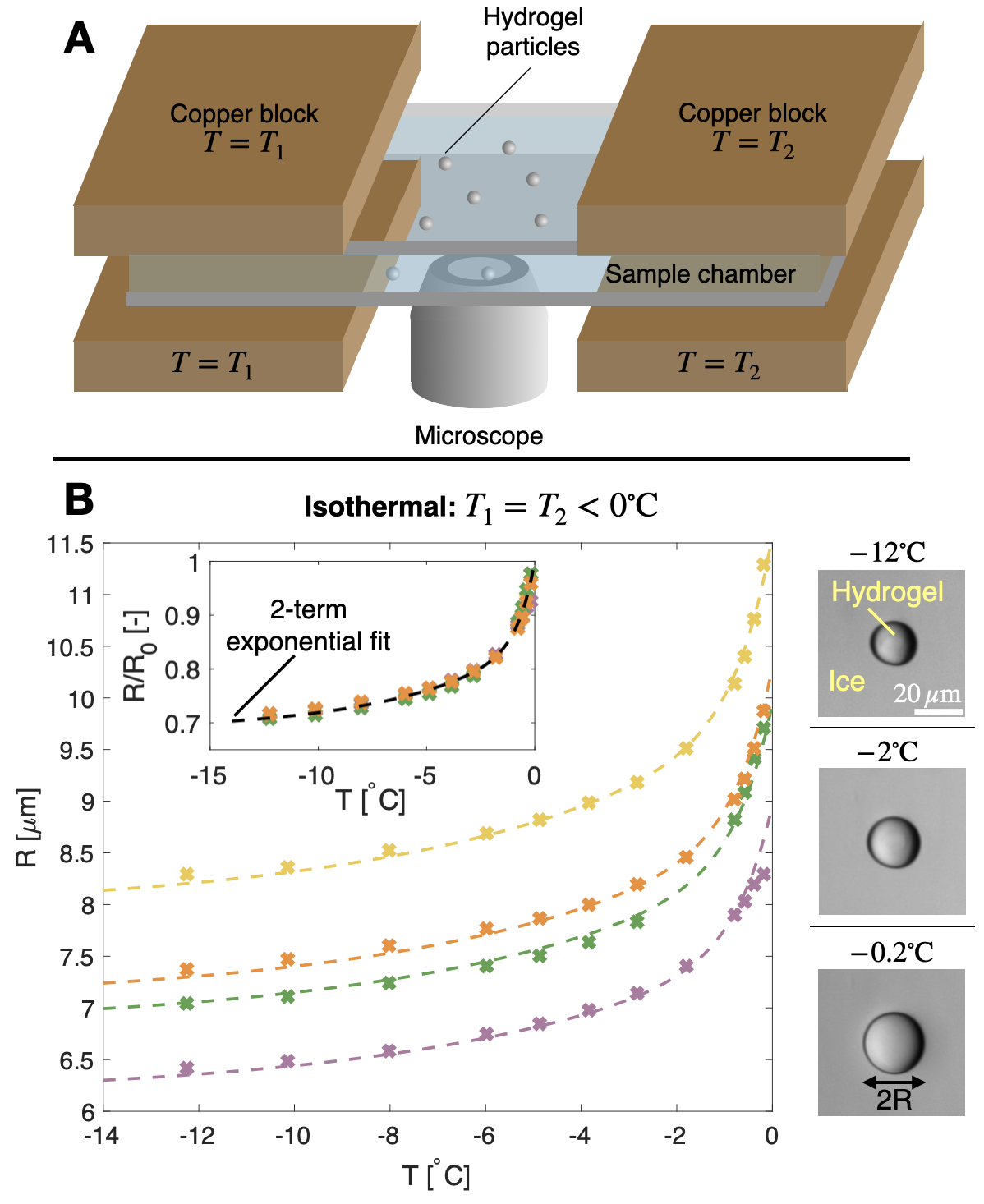}
    \caption{Observing the behavior of microscopic hydrogel particles embedded in ice. \textbf{A.} Schematic of the freezing apparatus.
    \textbf{B.} Main plot: how the radii of four different hydrogel particles in ice changes with temperature between -14 and 0$^\circ$C. At each temperature, particles are left for at least 5 minutes to ensure thermodynamic equilibrium is reached. Inset: all four data sets collapse when normalized by the radius of fully swollen particles, $R_0$. The dashed curve is equation (\ref{eqn:double_exp}). Images: a single particle at three different temperatures. }
    \label{fig:schem_isothermal}
\end{figure} 

We demonstrate the key physics involved in this technique by observing individual hydrogel particles in ice under isothermal conditions ($G=0$).
Figure \ref{fig:schem_isothermal}B shows a hydrogel particle, embedded in ice and equilibrated (by waiting at least 5 minutes) at three different temperatures.
The particle's radius, $R$, reduces significantly as the temperature decreases.
Indeed, we see this quantitatively when we plot $R(T)$ for four separate particles in the Figure.
In each case, $R$ reduces quickly by about 20\% as $T$ reduces from 0 to -2$^\circ C$, with a corresponding $\sim 50\%$ reduction in volume.
Subsequently, $R$ reduces more gradually with decreasing  temperature.

This shrinkage occurs due to a phenomenon commonly referred to as `cryosuction'.
Upon freezing, ice surrounds the particles, but does not grow into the hydrogel's porous mesh (prevented by capillarity via the Gibbs-Thomson effect \cite{webber2025cryosuction}).
Instead the ice sucks water out of the hydrogels, shrinking them by decreasing the hydrogel's liquid pore pressure, $p$.
This process is completely analogous to the process of a hydrogel shrinking in dry air.
The ice-induced shrinkage continues until the pore water reaches a thermodynamic equilibrium with the surrounding ice \cite{yang2024dehydration}, described by the generalized Clapeyron equation
(see Appendix) \cite{wettlaufer2006premelting,style2023generalized}:
\begin{equation}
p=P_i-\frac{\rho_w L (T_m-T)}{T_m}. 
\label{eqn:Clap}
\end{equation}
Here, $P_i$ is the ice pressure (at atmospheric pressure), while $\rho_w=1000\,\mathrm{kg}/\mathrm{m}^3$ is the density of water, $T_m=273\, \mathrm{K}$ is the freezing temperature of water at atmospheric pressure, and $L=3.34\times10^5\,\mathrm{J}/\mathrm{kg}$ is the specific latent heat of melting.
The equation fits with our observations, as $p$ decreases linearly with the undercooling $T_m-T$.
Thus, the colder it is, the larger the suction that is developed in a hydrogel, and the more the hydrogel will shrink (Figure \ref{fig:schem_isothermal}B).
Furthermore, the developed suctions are certainly enough to deform our hydrogels, as $p$ drops by about 1\,MPa per degree.
Thus, a few degrees of undercooling will significantly dehydrate the hydrogels (which have $O(100\,\mathrm{kPa})$ stiffness).

\begin{figure*}[hpbt]
    \centering
    \includegraphics[width=\linewidth]{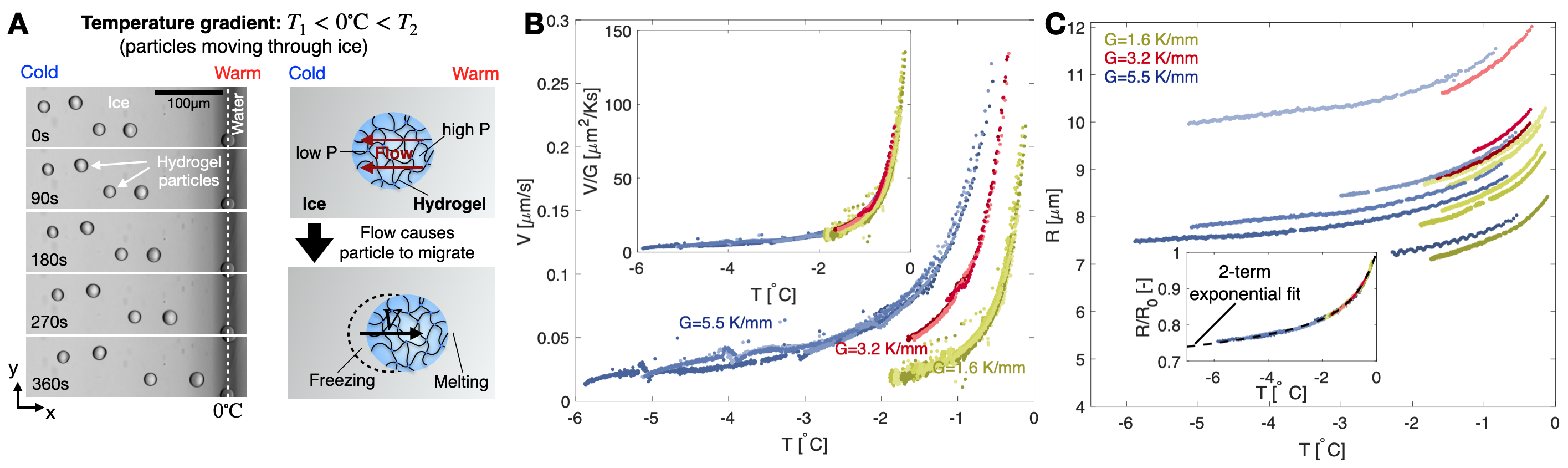}
    \caption{Hydrogels move through ice in a temperature gradient. 
    \textbf{A.} Images show PEGDA particles in ice in a temperature gradient ($G=3.2\,\mathrm{K/mm}$) at 90\,s intervals. Particles move to warmer temperatures while gradually swelling. The schematic shows the proposed mechanism: water melts on the warm side of a particle, flows through the particle (due to cryosuction), and refreezes on the cold side. This pushes the particle up the temperature gradient.
    \textbf{B.} The speeds of particles moving through ice in different temperature gradients. Different shades of colors denote separate particles. Inset: the data collapses when velocity is normalized by $G$.
    \textbf{C.} The radii of the same particles, as a function of their local temperature. Inset: the data collapses when normalized by the radius of of fully swollen particles, $R_0$. The dashed curve is equation (\ref{eqn:double_exp}).}
    \label{fig:gradient}
\end{figure*} 

Equation \ref{eqn:Clap} shows that we can shrink hydrogels simply by encapsulating them in ice, and reducing their temperature.
This shrinkage should be the same, regardless of the size of the hydrogel piece.
This is exactly what we measure: particles of different sizes, and bulk pieces of hydrogel all shrink by the same relative amount as they cool.
In our previous work, we measured the shrinkage of bulk layers of the same hydrogel \cite{feng2025characterizing}. There, the relative shrinkage as a function of temperature was empirically described by the double-exponential function:
\begin{equation}
R=R_0\left(A e^{\frac{T-T_m}{\Delta T_1}}+B e^{\frac{T-T_m}{\Delta T_2}} +C\right),
\label{eqn:double_exp}
\end{equation}
where $R_0$ is the radius of the swollen gel at 0$\,^\circ$C, and $A,B,C,\Delta T_1,\Delta T_2$ are fixed constants (see Appendix for details). This form also accurately captures how individual particle shrinkage, as shown by the dashed curves in Figure \ref{fig:schem_isothermal}B (fitted to obtain $R_0$ for each particle). 
Indeed, when we plot $R(T)/R_0$, the data all collapses onto a single curve (inset), confirming that cryosuction produces size-independent shrinkage.
Interestingly, the fact that bulk and microparticle hydrogels have the same shrinkage behavior also implies that these have very similar material properties, despite having different synthesis routes \cite{wyss2010capillary}.

\subsection{Hydrogel particles move through ice in a temperature gradient}

Hydrogels in ice also exhibit striking directional motion when they are also exposed to an applied temperature gradient. 
Figure \ref{fig:gradient}A shows a timelapse of a frozen particle suspension with $G=3.2\,\mathrm{K/mm}$ (see also SI Video 1). 
The particles move through the solid ice towards warmer temperatures, with velocities, $V\sim 0.1\, \mu\mathrm{m/s}$ along the temperature gradient (i.e. in the $x$ direction). 
There is no velocity in the perpendicular, $y$ direction (see SI).

This motion arises due to suction differences across the particle \cite{you2021thermal,rempel2001interfacial}.
Equation (\ref{eqn:Clap}) shows that $p$  is lower on the colder side of the particle than the warmer side.
This causes an induced pressure gradient, which pulls liquid through the particle towards the cold side (see Figure \ref{fig:gradient}A).
The flow is fed by ice melting on the warm side of the particle.
Simultaneously, it feeds new ice growth on the cold side, which pushes the particle in the opposite direction (towards the warm side) at the same rate that the water flows through the particle.
Indeed, when we prevent flow through the particles, by replacing hydrogels with impermeable particles, there is no observable particle movement (SI Video 2).

We obtain further confirmation of the movement mechanism by characterizing the precise particle velocities. 
Figure \ref{fig:gradient}B shows $V$ for 13 particles ($7\,\mu\mathrm{m}<R_0<13\,\mu\mathrm{m}$) in three different temperature gradients. For each different temperature gradient, all the data collapses onto a single curve (blue: $G=5.5\, \mathrm{K}/\mathrm{mm}$, red: $G=3.2\, \mathrm{K}/\mathrm{mm}$, and green: $G=1.6\, \mathrm{K}/\mathrm{mm}$). Thus, $V$ only depends on $T$ and $G$ and is independent of $R_0$. The $T$ dependence is strong, with a sharp upturn in speed as particles reach temperatures $\gtrsim -1\,^\circ$C. The vertical shift of curves with increasing $G$ shows that increasing $G$ also increases $V$.
Indeed, when we plot the normalized velocity $V/G$ against $T$, all the data collapses onto a single curve, suggesting that $V\propto G$.

These observations match the proposed mechanism. 
As particle motion is driven by flow through the hydrogel, we expect particle velocity to be given by Darcy’s law (\ref{eqn:darcy}): $V=-\frac{k}{\eta} |\nabla p|$. 
Furthermore, $p$ depends linearly on temperature (equation (\ref{eqn:Clap})), so that $\nabla p\propto G$, and $V\propto k G$.
This last result explains three features of our data. 
Firstly, it predicts that $V$ is independent of $R_0$.
Secondly, it confirms that $V$ is linearly proportional to $G$.
Finally, it suggests that $V$ increases with temperature because $k$ also increases strongly with temperature.
This increase in $k$ with temperature is expected, due to particles swelling, causing their pore size to grow.
Most importantly, the fact that $V\propto kG$ gives us a direct way to obtain values of $k$ by simply measuring $V$.

Before using our data to extract material properties, we note that the dynamic experiments also give us the same information about swelling equilibrium that we previously obtained from isothermal freezing experiments (Figure \ref{fig:schem_isothermal}B).
However, the dynamic data gives a much greater density of data points.
In Figure \ref{fig:gradient}C, we show plots of $R(T)$ for moving particles.
As before, each plot is well described by equation (\ref{eqn:double_exp}), allowing us to fit each particle's swollen $R_0$.
When we plot $R(T)/R_0$, all the data collapses onto the single curve (inset, Figure \ref{fig:gradient}C) -- just as we saw previously in Figure \ref{fig:schem_isothermal}B.

\subsection{Extracting material properties from hydrogel freezing behavior}

The results above allow us to extract hydrogel properties as a function of swelling (or equivalently, as a function of $\phi_p$).
Here, we first show how to use swelling data to measure compressional properties such as effective pressure and drained compressional modulus.
Second, we use velocity measurements to obtain hydrogel permeability.
Finally, we can combine the measurements together to obtain the poroelastic diffusivity.

\begin{figure}[hpbt]
    \centering
    \includegraphics[width=0.9\linewidth]{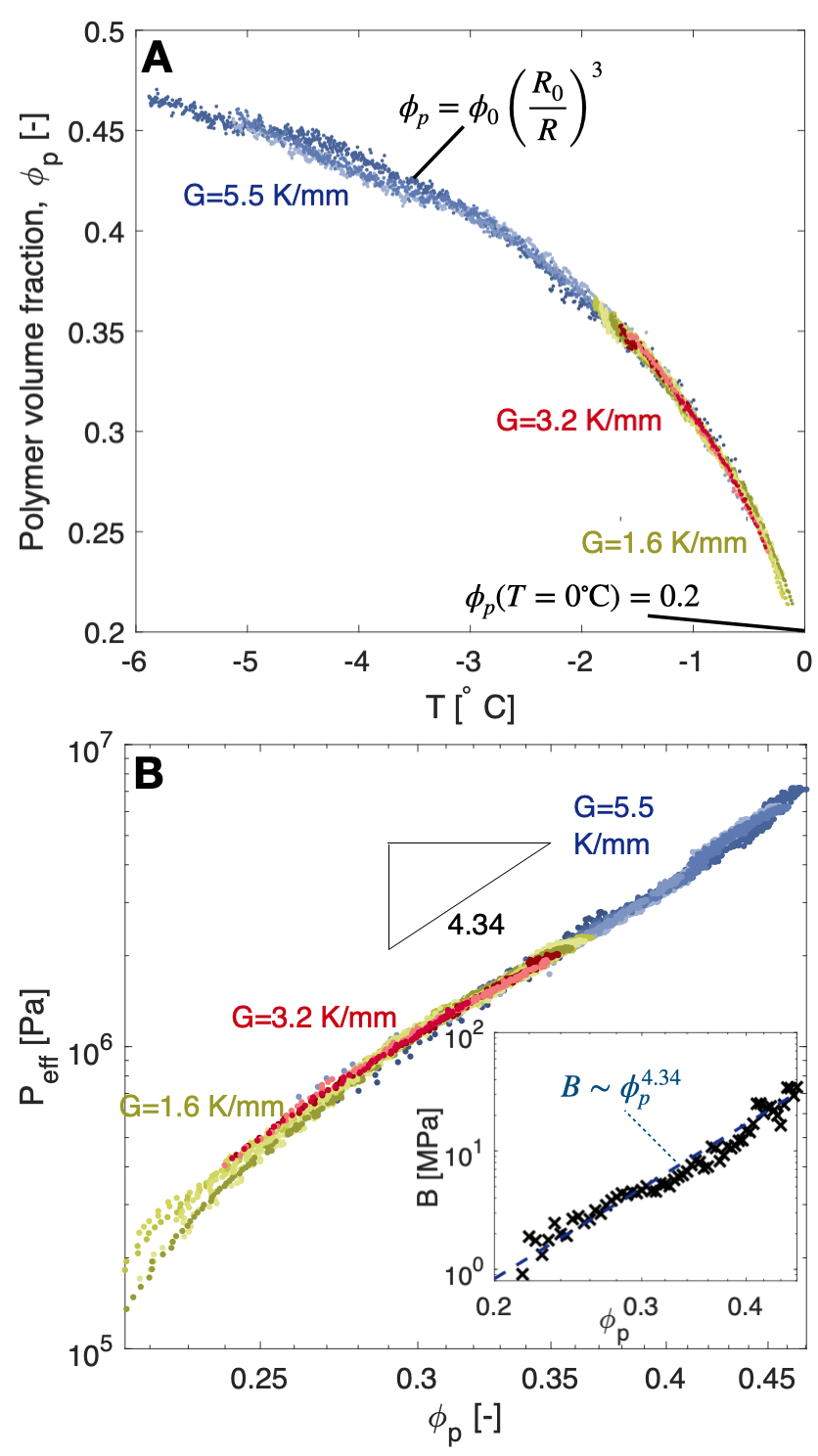}
    \caption{The compressional properties of the hydrogel particles observed in Figure \ref{fig:gradient}.
    \textbf{A.} The polymer volume fraction as a function of local temperature, calculated from relative changes in radii.  
    \textbf{B.} The effective pressure as a function of polymer fraction. Inset: the bulk modulus, $B=\phi_p\frac{dP_{\mathrm{eff}}}{d\phi_p}$.}
    \label{fig:compression}
\end{figure}

To obtain properties as a function of polymer content, we need to determine $\phi_p$.
We calculate this, using the fact that the volume of polymer in particles is fixed, so that $\phi_p R^3=\phi_p^0 R_0^3$, where $\phi_p^0=20\%$ is $\phi_p(0\,^\circ$C).
Thus, we can use measurements of $R(T)/R_0$ (Figure \ref{fig:gradient}C) to obtain $\phi_p(T)$.
The results are shown in Figure \ref{fig:compression}A, and effectively give a master curve for how hydrogels dehydrate as cryosuction increases.

Next, we calculate the effective pressure, $P_{\mathrm{eff}}$ in the hydrogel, and how it changes with polymer content.
To explain $P_{\mathrm{eff}}$, we note how the total pressure in a hydrogel is split into different components \cite{feng2025characterizing,louf2021under,bertrand2016dynamics}: 
\begin{equation}
P_{tot}=P_{mix}+P_{el}+p.
\label{eqn:P_components}
\end{equation}
The mixing pressure, $P_{mix}$, comes from attractive polymer/solvent interactions; the elastic pressure, $P_{el}$, comes from the elasticity of the polymer network in the gel; and $p$ is the pore pressure.
$P_{tot}$ and $p$ are set by external conditions (respectively the mechanical pressure on the outside of the gel and the applied suction).
However, $P_{\mathrm{eff}}(\phi_p)=P_{mix}+P_{el}$ is a function of the current conformation of the network, and thus tells us the mechanical response of the gel at a given swelling state.
For example, in a drained experiment (immersed in water, with $p=0$), $P_{\mathrm{eff}}=P_{tot}$ is the mechanical pressure needed to compress a gel to a volume fraction $\phi_p$.
Alternatively, when a hydrogel dries in air at atmospheric pressure ($P_{tot}=0$), $P_{\mathrm{eff}}=-p$: the suction acts to pull any available water into the gel. 

We obtain $P_{\mathrm{eff}}(\phi_p)$ by combining equations (\ref{eqn:Clap}) and (\ref{eqn:P_components}) to find that $P_{\mathrm{eff}}=\frac{\rho_w L_m (T_m-T)}{T_m}$.
Then, we combine this expression with the data for $\phi_p(T)$ from Figure \ref{fig:compression}A to eliminate $T$.
The results in Figure \ref{fig:compression}B show that $P_{\mathrm{eff}}$ increases strongly with increasing $\phi_p$: by almost two orders of magnitude over the measured range.
This increase is consistent with previous measurements, and is likely due to the fact that $P_{\mathrm{eff}}\approx P_{mix}$ at higher concentrations \cite{correas2026hydrogel}.
$P_{mix}$ is a strong, nonlinear function of $\phi_p$ for hydrophilic polymers, often well described by a power-law function.
Indeed, here, $P_{\mathrm{eff}}\sim \phi^{4.3}$.

Knowing $P_{\mathrm{eff}}$ allows us to measure the drained compressional modulus, $B(\phi_p)$ -- which measures how hard it is to squeeze water out of a gel in a drained test.
By definition, $B=-V\frac{\partial P_{tot}}{\partial V}$.
In a drained test,  $P_{tot}=P_{\mathrm{eff}}$, and $V\propto 1/\phi$, so $B= \phi_p\frac{\partial P_{\mathrm{eff}}}{\partial \phi_p}$.
We calculate this using the data in Figure \ref{fig:compression}B, and give the results in the inset.
$B\sim1\,$MPa when the gel is swollen, but increases dramatically as the hydrogel dries out, reaching $\sim 30\,$MPa when $\phi_p= 45\,\%$.
In the intervening range of $\phi_p$, we again see approximate power-law behavior:  
$B\sim \phi^{4.3}$.

\begin{figure}[hpbt]
    \centering
    \includegraphics[width=0.9\linewidth]{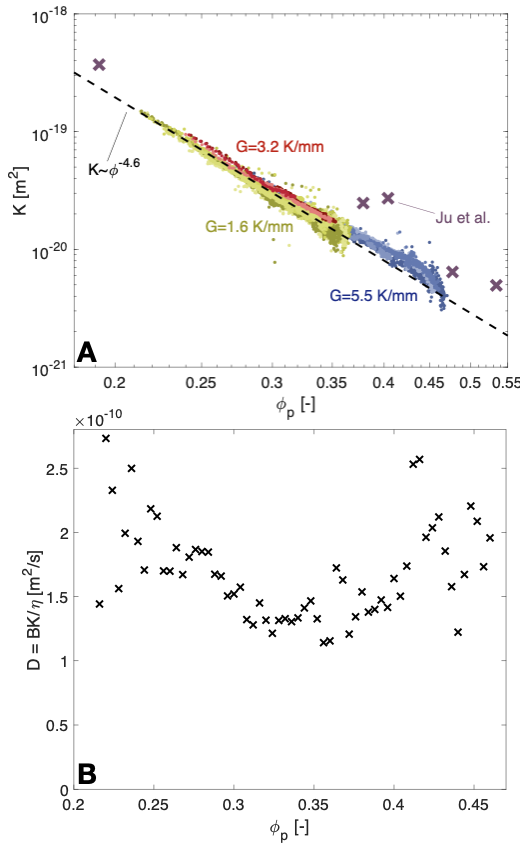}
    \caption{Transport properties of water in the hydrogel particles observed in Figure \ref{fig:gradient}.
    \textbf{A.} Hydrogel permeability. Crosses show measurements in similar hydrogels by Ju et al. \cite{ju2010characterization}. Dashed line shows a best fit power law.
    \textbf{B.} Poroelastic diffusivity.}
    \label{fig:permeability}
\end{figure}

Next, we extract measurements of $k(\phi)$ from the speed that hydrogel particles move through ice.
Following You et al. \cite{you2021thermal} we combine Darcy's law (\ref{eqn:darcy}) with equation (\ref{eqn:Clap}) to find that $|\nabla p|=-\rho_w L_m G_h/T_m$.
The relevant temperature gradient here is that inside the hydrogel particle, $G_h=3\lambda_i G /(2\lambda_i+\lambda_h)$ (see Appendix), which differs from $G$ due to differences in thermal conductivity in the ice ($\lambda_i$) and in the gel ($\lambda_h$).
Combining everything together yields
\begin{equation}
k=\frac{\eta T_m (2\lambda_i+\lambda_h)}{3 \rho_w L_m \lambda_i} \frac{V}{G}
\label{eqn:perm}
\end{equation}
(see the Appendix for a full derivation).
To apply this equation, we use  $\lambda_h=\phi_p\lambda_p+(1-\phi_p)\lambda_w$, where $\lambda_p=0.23\,$W/(mK), and $\lambda_w$ is the thermal conductivity of water. 
Furthermore, we use reference correlations for the values of $\eta(T)$ \cite{kestin1978viscosity}, $\lambda_w(T)$ \cite{ramires1995standard}, and $\lambda_i(T)$ \cite{cuffey2010physics}. 
Then, inserting the data from the inset of Figure \ref{fig:gradient}B and Figure \ref{fig:compression}A gives $k(\phi_p)$. 
The results are given in Figure \ref{fig:permeability}A, and show how permeability drops dramatically with $\phi_p$. Indeed, $k$ reduces by almost two orders of magnitude as $\phi_p$ increases from 20\% to 45\%. This strong decay in $k(\phi_p)$ is well described by a power law over this range, with $k\sim \phi_p^{-4.6}$ (see dashed line).

Finally, we obtain the poroelastic (or cooperative) diffusivity, $D$. 
This parameter dictates how fast water swelling/de-swelling fronts move through hydrogels \cite{hong2008theory,hu2010using}. 
$D=\frac{k}{\eta}\left(B+\frac{4}{3}\mu\right)$,
where $\mu$ is the gel’s shear modulus \cite{fujiyabu2018three,doi2009gel}. 
However, we can simplify this, as for our gels, $\mu\approx 300\,\mathrm{kPa} \ll B$ \cite{correas2026hydrogel}, so that $D\approx kB/\eta$.
We calculate this by  combining our measurements of $B$ and $k$, and using the viscosity of water at room temperature.
The results are shown in Figure \ref{fig:permeability}B, and show that $D\approx 1.5\times 10^{-10}\,\mathrm{m}^2/\mathrm{s}$ is rather constant as the gel shrinks.

\subsection{Comparison with literature results}

Our technique gives results that are consistent with, but much higher-resolution than literature results from conventional techniques.
For example, in Figure \ref{fig:permeability}A, we compare our results with the sparse available literature data for $k$ for similar PEGDA hydrogels.
To obtain this data, Ju \emph{et al.} \cite{ju2010characterization} prepared hydrogel membranes with a range of different polymer contents (by varying the membranes' as-prepared polymer content).
Then, they extracted $k$ by measuring water flow rates through these membranes in response to externally-applied pressure gradients.
The results are given as crosses in Figure \ref{fig:permeability}A, and show essentially the same trend in $k(\phi_p)$ as our data, with only an $O(1)$ factor difference.
This is very reasonable agreement -- especially given that Ju et al. used a different synthesis protocol, different photoinitiators, and different $\phi_p^0$ when preparing their gels.
For the other measured parameters ($P_{\mathrm{eff}}(\phi_p)$, $B(\phi_p)$ and $D(\phi_p$)), it is challenging to compare our results to literature data.
This is because there is very little data for these parameters due to the lack of appropriate measurement techniques.
However, we note that our measurements of $D$ are in the middle of the typical range of poroelastic diffusivities that are seen  in synthetic, covalently-bonded hydrogels ($10^{-11}\,\mathrm{m}^2\lesssim D \lesssim 10^{-9}\,\mathrm{m}^2$ \cite{caccavo2018hydrogels,raasmark2005fast,fujiyabu2018three}).

The high resolution of our data sets allow us to test classical predictions hydrogel transport properties (e.g. \cite{sakai2020physics}).
For example, a widely used assumption is that a hydrogel essentially behaves as a loosely-crosslinked, semi-dilute polymer solution.
In such polymer solutions, the polymers coil up into necklace-like chains of spherical blobs of size $R_b$ \cite{de1979scaling}.
This blob size sets the hydrogel's `pore size'.
If we assume that pressure gradients across the gel drive hydrodynamic flows through the hydrogel pores, then $k\sim R_b^2$ \cite{offeddu2018relationship}.
Furthermore, the scaling for osmotic pressure in a semi-dilute polymer solutions gives that  $P_{\mathrm{eff}}\sim B\sim k_bT/R_b^3$, where $k_b$ is Boltzmann's constant \cite{des1975lagrangian,yasuda2020universal}. 
For a polymer in a good solvent (like a hydrogel network in water), $R_b\sim \phi_p^{-3/4}$ \cite{rubinstein2003polymer}. 
Thus, we would expect $k\sim \phi_p^{-1.5}$ \cite{fujiki2016friction}, $P_{\mathrm{eff}}\sim B\sim \phi_p^{2.25}$ \cite{feng2025characterizing}, and $D\sim\phi_p^{0.75}$ \cite{fujiyabu2018three}.

These predictions are markedly different to our measurements ($k\sim \phi_p^{-4.6}$, $P_{\mathrm{eff}}\sim B\sim \phi_p^{4.3}$, and $D\approx \mathrm{constant}$).
This discrepancy could arise for several reasons.
Firstly, our PEGDA gels are made with relatively short oligomers (13 monomer units), capped with two relatively hydrophobic acrylate groups.
Thus, it is questionable whether the chains between crosslinks are long enough to form blobs \cite{rubinstein2003polymer}.
Secondly, our gels are likely rather inhomogeneous: previous work has shown that PEGDA gels have a mixture of large and small pores \cite{feng2025controlling,malo2015heterogeneity}.
When these hydrogels are fully swollen, the presence of large pores will make the gels permeable and soft.
However, when the gels are dehydrated, the larger pores will collapse, leaving much smaller average pore sizes.
This should dramatically reduce permeability and increase stiffness -- much more than would be expected in a hydrogel with an ideal polymer network.
Indeed, recent work has proposed that characterizing such changes is a sensitive technique for measuring network inhomogeneity \cite{ito2026hierarchical}.

While the predictions above assume hydrodynamic flows through hydrogel pores, an
alternative viewpoint may better explain the data.
In this viewpoint, water transport through hydrogels is a diffusive process -- governed by a diffusivity $D$ \cite{hegde2022two}.
This diffusion should be very similar to how water diffuses through uncrosslinked polymer solutions (with diffusivity, $D_s$), implying that $D\approx D_s$.
Previous measurements have shown that $D_s \sim 10^{-10}-10^{-9}\,\mathrm{m}^2/\mathrm{s}$ in uncrosslinked PEG solutions, and is a surprisingly weak function of polymer content -- only decreasing by about one order of magnitude as $\phi_p$ increases from 0-100\% \cite{moon2024nanoscale}.
This is indeed similar behavior to what we  see in our experiments  (Figure \ref{fig:permeability}B).
The constraint that $D$ is a weak function of $\phi_p$ then also explains the similar size of the scalings of $B$ and $k^{-1}$ (as $B\sim D/k$).

\subsection{Conclusions}

In conclusion, we have demonstrated a technique for precisely characterizing the properties that control water transport in hydrogels -- as a function of polymer content.
This technique works by simply observing the speed and size of hydrogel particles moving through ice in a temperature gradient.
These observations are then converted into measurements of permeability, compressibility, effective pressure, and poroelastic diffusivity.

Our technique has several key advantages over existing alternatives.
Firstly, while typical macroscopic measurement techniques need $\gtrsim O(\mu\mathrm{L})$ volumes, our approach only requires much smaller, $O(\mathrm{pL})$ volumes of hydrogels.
This removes the requirement of obtaining homogeneous, macroscopic samples of (often expensive) hydrogels for testing.
Secondly, most macroscopic measurement techniques are limited by the fact that they can only characterize fully-swollen gels.
This makes it very challenging to measure how properties change with polymer content.
By contrast, our technique allows us to measure properties at a range of different $\phi_p$ -- on a single hydrogel piece.
Finally, standard approaches are limited to a single data point per experiment, giving relatively sparse plots for each material property.
Our approach can give orders of magnitude higher resolution, with thousands of data points from a single experiment.
This allows the precise testing of long-standing scaling laws and theories.

We envisage several directions for future work.
Firstly, we can significantly expand the scope of the technique.
For example, it should not only work with perfectly spherical hydrogel particles.
We anticipate that we can also use small pieces of bulk hydrogels (e.g. broken up via cryo-milling \cite{yuan2026hydrogel}) to measure the same properties.
This would allow the direct characterization of bulk samples (e.g. biological materials).
Furthermore, our technique should also be applicable for characterizing other types of materials, like polymer solutions and nanoporous solids.
In the latter case, liquid solution droplets can be easily encapsulated in ice, and will behave in a similar way to our hydrogel particles.
Tracking droplet size and speed should then allow one to measure properties such as solute diffusivity and osmotic pressure.

Beyond this, our results highlight the need to develop new theories that can accurately describe hydrogel transport behavior.
Here, we have shown how commonly-used scaling laws for transport properties do a very poor job of describing our results -- despite their success in modeling gels with ideal, homogeneous networks (e.g. tetra-PEG gels \cite{sakai2008design}).
This suggests that these scaling laws may need to be adapted to model `real' hydrogels with naturally occurring inhomogeneities.
We anticipate that this process requires a precise  characterization of a wide range of different hydrogels -- something which we can now achieve with this technique.

\section*{Materials and Methods}

\subsection*{Synthesis of hydrogel particles}

We synthesize poly(ethylene glycol) diacrylate (PEGDA) particles using droplet microfluidics, using a flow-focusing configuration~\cite{jelken2022tuning, shen2026programmable}.
Microfluidic devices are fabricated by soft lithography. The microfluidic channel patterns were designed using L-Edit software (Siemens, Germany), as illustrated in Figure~S1. The patterns are transferred onto a 5-inch chrome-coated glass mask using a DWL-66 mask writer (Heidelberg Instruments, Germany), followed by standard mask development. The resulting photomask is used to pattern an approximately \(15~\mu\mathrm{m}\)-thick layer of SU-8 3025 photoresist (MicroChem, USA) on a 4-inch silicon wafer. The SU-8 photoresist is deposited according to the manufacturer's recommended protocol by spin-coating at \(1500~\mathrm{rpm}\) for \(40~\mathrm{s}\), with an acceleration of \(300~\mathrm{rpm\,s^{-1}}\).

PDMS replicas are prepared using Sylgard 184 (Dow Corning, USA) by mixing the PDMS base and curing agent at a ratio of 10:1. The mixture is poured over the patterned silicon master, and trapped air bubbles are removed under vacuum for \(30~\mathrm{min}\). The PDMS is subsequently cured at \(65~^\circ\mathrm{C}\) for approximately \(12~\mathrm{h}\). After curing, the PDMS replicas are carefully peeled from the silicon master and rinsed three times with isopropanol, followed by distilled water.

The PDMS replicas and glass slides are activated by oxygen-plasma treatment at \(100~\mathrm{W}\) for \(40~\mathrm{s}\). Immediately after plasma treatment, the activated surfaces are brought into contact to form covalent siloxane bonds through the condensation of surface silanol groups. Finally, the assembled devices are heated at \(120~^\circ\mathrm{C}\) for \(5~\mathrm{min}\) to strengthen the bond and remove residual water.

To prepare a PEGDA precursor solution, we mix \(200~\mu\mathrm{L}\) of PEGDA oligomer (\(M_\mathrm{n} = 700~\mathrm{g\,mol^{-1}}\)) with \(800~\mu\mathrm{L}\) of Milli-Q water, and then add Lithium phenyl-2,4,6-trimethylbenzoylphosphinate (LAP) photoinitiator at a final concentration of \(10~\mathrm{mg\,mL^{-1}}\).
This solution is then transferred into a glass syringe.

We generate water-in-oil droplets at the cross-junction of the PDMS microfluidic device.
The aqueous precursor solution serves as the dispersed phase, while HFE-7500 fluorocarbon oil containing \(1.8~\mathrm{vol\%}\) Krytox FSH is used as the continuous phase.
Droplet size is controlled by systematically varying the relative flow rates of the dispersed and continuous phases. The aqueous phase is injected at \(30\)--\(60~\mu\mathrm{L\,h^{-1}}\), whereas the oil phase is supplied at either \(120\) or \(180~\mu\mathrm{L\,h^{-1}}\), corresponding to aqueous-to-oil flow-rate ratios of \(0.17\)--\(0.50\).
The resulting emulsions are collected in \(2~\mathrm{mL}\) vials over \(4~\mathrm{h}\). Then, we initiate photopolymerization by irradiating the emulsions at \(365~\mathrm{nm}\) for \(10~\mathrm{min}\) using a \(6~\mathrm{W}\) (an illumination intensity of \(0.75~\mathrm{mW\,cm^{-2}}\)) UV light source. 
We then store the emulsions overnight at room temperature.

The following day, we remove the bulk fluorocarbon oil, and sequentially wash the resulting microgels with HFE-7500, hexane, 1,4-dioxane, isopropanol, and Milli-Q water.
During each washing step, we allow the microgels to separate from the liquid phase by sedimentation or flotation, as appropriate, after which we carefully remove and replaced the liquid phase with fresh solvent.
Each washing step is repeated three times.
We perform HFE-7500, hexane, and 1,4-dioxane washing using gravity-driven separation, isopropanol washing  using centrifugation at \(8{,}000~\mathrm{rpm}\) for \(5~\mathrm{min}\),  and  Milli-Q water washing using  centrifugation at \(10{,}000~\mathrm{rpm}\) for \(10~\mathrm{min}\).
Finally, we redisperse the purified microgels in fresh Milli-Q water.

\subsection*{Freezing experiments}

The sample cell consists of a $50 \times 25\,\text{ mm}$ glass slide, onto which a SecureSeal$^{\mathrm{TM}}$ imaging spacer is affixed to form a well. 
Approximately 100 $\mu$L of the purified microgel suspension is pipetted into the well, which is then sealed with a second $50 \times 25\,\text{ mm}$ glass slide. 
The assembled cell is inserted into a custom-built temperature-gradient freezing stage.

To form ice in the freezing cell, we first cool one side of the sample slightly below
0\textdegree{}C ($T_1 < 0^\circ$C). 
This is achieved by independently tuning the temperatures of the copper blocks in our freezing stage. 
Then, we nucleate ice by touching the cold side of the copper block with a cotton swab that has been dipped in liquid nitrogen.
Following nucleation, the temperature of the two sides is tuned to the desired temperature gradient, and the system is equilibrated until a stable temperature gradient is established across the sample for the freezing experiment.

Three-dimensional (3D) image stacks of the hydrogel particles are acquired using a Nikon Ti2 Eclipse microscope equipped with a 3i spinning disk confocal system and a 20x air objective (NA 0.17).
The imaging protocol begins by identifying particles of interest within the temperature gradient (ones that are well separated from other particles in the ice). 
A time-lapse series of brightfield 3D image stacks (z-step: 1 $\mathrm{\mu m}$; total z-range: 15-30 $\mathrm{\mu m}$) is acquired at intervals of 30-60 seconds. 
To track particles as they migrate under a thermal gradient, the microscope stage is occasionally repositioned toward warmer regions between acquisitions. 
This process is continued until the particles' incorporation by the ice-water interface is fully captured.

Particle radii are determined from the 3D stacks using a custom MATLAB script. 
The script first identifies the plane of best focus and then detects its perimeter to calculate the radius.

\subsection*{Appendix: Cryosuction theory}
The generalized Clapeyron equation relates the pressures of ice and water ($P_i$ and $p$ respectively) that are in equilibrium at the same temperature $T$ \cite{style2023generalized}:
\begin{equation}
    \frac{P_i-P_{atm}}{\rho_i}-\frac{p-P_{atm}}{\rho_w}=L\frac{T_m-T}{T_m}.
\end{equation}
Here, $P_{atm}$ is atmospheric pressure, $L$ is the specific latent heat of melting, and $\rho_i,\rho_w$ are the densities of ice and water, respectively. $T_m$ is the melting temperature of bulk ice at atmospheric pressure.
Here, ice is stress-free and exposed to the atmosphere, so $P_i=P_{atm}$.
Inserting this expression and rearranging gives 
\begin{equation}
    p=P_{atm}-\frac{\rho_w L (T_m-T)}{T_m}.
    \label{eqn:clap_supp}
\end{equation}

To calculate permeability, we need to know the pressure gradient inside hydrogel particles.
To obtain this, we first show that $p=p(T)$ inside the particles.
We start by using Darcy's law for the flux of liquid inside the particles: $\mathbf{Q}=-\frac{k}{\eta} \mathbf{\nabla p}$.
We note that $-|\mathbf{Q}|=V$ is just the moving velocity of the particles, and is thus a constant (assuming particles are small enough that their properties do not vary across their width).
Thus, $\nabla.\mathbf{Q}\propto \nabla^2 p =0$.
Furthermore, the temperature in the slowly moving particle satisfies $\nabla^2 T=0$, and thermodynamic equilibrium between ice and water requires that equation (\ref{eqn:clap_supp}) holds on the particle/ice boundary.
In fact, a general solution to these equations with the boundary condition is that equation (\ref{eqn:clap_supp}) holds in the entire interior of the particle.
Thus, the pressure gradient inside the hydrogel particle comes from the gradient of equation (\ref{eqn:clap_supp}):
\begin{equation}
    |\nabla p| = \frac{\rho_w L}{T_m}G_p,
\end{equation}
where $G_p$ is the temperature gradient inside the particle.

We obtain $G_p$ from the work of You et al. \cite{you2021thermal}.
They have shown that
\begin{equation}
    G_p=\frac{3\lambda_i}{2\lambda_i+\lambda_p}G-\frac{\rho_wLV}{2\lambda_i+\lambda_p},
    \label{eqn:you}
\end{equation}
where $\lambda_i,\lambda_p$ are the thermal conductivities of ice and particle respectively.
The first term accounts for differences between $G$ and $G_p$ due to differences in thermal conductivity. The second term accounts for the fact that there is melting on the warm side of the particle, and refreezing on the cold side.
The associated removal/release of latent heat makes a small change to the temperature gradient.
In our experiments, values of $3\lambda_i G$ are at least two orders of magnitude bigger than $\rho_w L V$, so we ignore the latent heat term.

Finally, we can combine the results above to obtain an expression relating particle speed to the permeability.
In particular, we combine Darcy's law with equations (\ref{eqn:clap_supp},\ref{eqn:you}) to obtain equation (\ref{eqn:perm}).

Note that this result assumes that ice crystallization kinetics plays a minimal role. This is expected, as ice growth rates here are very slow.
We can confirm that kinetic effects are minimal by comparing the magnitude of temperature jumps at the ice/gel interface due to kinetics, $\Delta T_i$, with temperature changes across particles, $\Delta T_p$ (the latter of which drives cryosuction).
If $\Delta T_i\ll \Delta T_p$, then kinetic effects are minimal.
We estimate $\Delta T_i$ by using the kinetic coefficient for solidification, $\Gamma\sim 10^{-2}\,\mathrm{m/sK}$ \cite{addula2022kinetic}.
This relates $\Delta T_i$ to  the growth speed of the ice/water interface: $\Delta T_i=V/\Gamma$.
For typical values of $V\sim 10^{-7}\,\mathrm{m/s}$, we find that $\Delta T_i\sim 10^{-5}\,$K.
By comparison, the temperature difference across particles $\Delta T_p=GR\sim 10^{-2}\,$K.
Thus, indeed $\Delta T_i\ll\Delta T_p$, and kinetic effects are negligible.

\subsection*{Appendix: Deswelling of bulk hydrogel by cryosuction}
\begin{figure}[hpbt]
    \centering
    \includegraphics[width=0.8\linewidth]{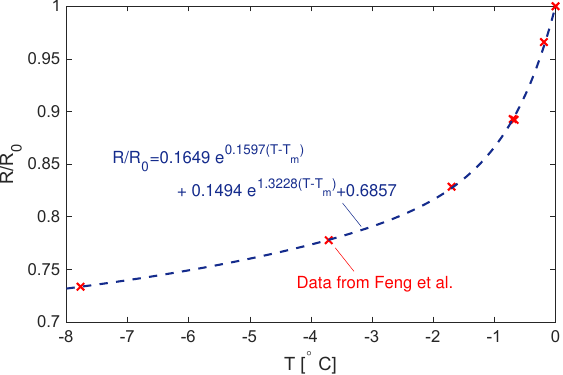}
    \caption{Data for the equilibrium swelling of bulk hydrogel slabs in contact with ice \cite{feng2025characterizing}. This is well described by the double exponential equation (\ref{eqn:double_exp}).}
    \label{fig:fitting_bulk}
\end{figure}

We can compare the equilibrium size of our hydrogel particles in ice to data from bulk hydrogels from Feng et al. \cite{feng2025characterizing}.
They measured the relative volume changes, ${\cal{V/V}}_0$ of bulk pieces of PEGDA hydrogel (from the same precursor solution) in contact with ice, as the temperature changed.
To make a direct comparison, we calculate the effective radius change of the hydrogel: $R/R_0=({\cal{V/V}}_0)^{1/3}$, and plot this in Figure \ref{fig:fitting_bulk}.
This data is well fit by the double exponential equation (\ref{eqn:double_exp}), with $A=0.1649$, $B=0.1494$, $C=0.6857$, $\Delta T_1=6.2617$, and $\Delta T_2=0.7560$.

\begin{acknowledgments}

YF and RWS acknowledge support from the Swiss National Science Foundation (200021-212066). We acknowledge helpful conversations with Grae Worster, Shaohua Yang and Sylvain Deville.

\end{acknowledgments}

\appendix

%



\end{document}